\documentclass[twocolumn,superscriptaddress]{revtex4}

\usepackage{epsfig}
\usepackage{latexsym}
\usepackage{bm}

\newcommand{\be}{\begin{equation}}
\newcommand{\ee}{\end{equation}}
\newcommand{\bea}{\begin{eqnarray}}
\newcommand{\eea}{\end{eqnarray}}

\def\beginwide{
        \end{multicols} \vspace*{-0.5cm} \noindent
        \rule{3.5in}{.1mm}\rule{.1mm}{5mm} \widetext \medskip }
\def\beginwidetop{
        \end{multicols} \vspace*{-0.5cm} \noindent
        \widetext \medskip }
\def\endwide{
        \hspace*{3.35in}~\rule[-5mm]{.1mm}{5mm}\rule{3.5in}{.1mm}
        \begin{multicols}{2} \vspace*{-1.0cm} \noindent }
\def\endwidebottom{
        \begin{multicols}{2} \vspace*{-1.0cm} \noindent }

\begin{document}
\title{Driven Diffusive Systems with Disorder} 
\author{Mustansir Barma} 
\affiliation{Department of Theoretical Physics, 
Tata Institute of Fundamental Research, Mumbai 400005, India} 
\date{\today}
\begin{abstract}
We discuss recent work on the static and dynamical properties of the asymmetric
exclusion process, generalized to include the effect of disorder. We
study in turn: random disorder in the properties of particles; disorder in 
the spatial distribution of transition rates, both with a single easy direction
and with random reversals of the easy direction; dynamical disorder, where 
particles move in a disordered landscape which itself evolves in time.
In every case, the system exhibits phase separation; in some cases, it is 
of an unusual sort.  The time-dependent
properties of density fluctuations are in accord with the
kinematic wave criterion
that the dynamical universality class is unaffected
by disorder if the kinematic wave velocity is nonzero.
\end{abstract}
\maketitle

\section{Introduction}

Simple models of particles moving on a lattice have shed a good deal
of light on the general question of collective effects in transport.
Even the simplest sorts of interactions between particles can induce
collective phenomena involving large numbers of particles, ranging from
the motion of density fluctuations as a
kinematic wave, to the occurrence of phase transitions to states with
macroscopically inhomogeneous density (jams).  Models which display
these phenomena include the asymmetric simple exclusion process
(ASEP), the simplest model which incorporates directed motion and
mutual exclusion \cite {spitzer,stinch,schutzreview}, and its generalizations 
which include more realistic features which mimic vehicular traffic
\cite{traffic}.

Here we are concerned with the effects of extensive disorder on
interacting-particle transport in one dimension.  As in equilibrium
systems, disorder has a strong effect on the properties of driven
systems which reach a nonequilibrium steady state.  However,
in contrast to equilibrium systems, relatively little is known, in a
general way, about the effects of quenched disorder on nonequilibrium
systems, as the absence of detailed balance, together with the
breaking of translational invariance, makes even the determination of the
steady state weights difficult in general.  In this backdrop,
the study of simple models of
disordered nonequilibrium systems has proved valuable, as it
reveals how 
interactions, drive and disorder combine to produce new types of states;
the simplicity of the models allows for a detailed characterization
and understanding.

In this paper, we will review some results obtained for a disordered
system of particles moving stochastically on a one-dimensional lattice.  
We restrict ourselves to the disordered ASEP, in which interactions
between particles enter through rules for the hopping rates -- the
hard-core constraint is modelled by forbidding any move which would
lead to more than one particle per site.  We will consider three
types of disorder.  The first is
{\it particle-wise} quenched disorder, in which different
particles have different properties which do not change in time.
Secondly, we will consider {\it space-wise} quenched disorder, in
which particles do not differ from each other, but are subject to
hopping rates which are randomly distributed in space.  Finally, we
will turn to {\it dynamical disorder} where particles move in a
disorder-induced landscape, which is itself evolving in time. 
Interestingly, there are inter-relations
between the effects of the three types of disorder, which we 
discuss below.

It is clear that extensive disorder at the microscopic scale would induce
inhomogeneities in the particle density on a similar scale. 
Furthermore, disorder can also induce variations of the density on a
{\it macroscopic} scale, associated with phase separation \cite{krug} .
The models considered here share this feature, but in some cases the
phase separated state has quite unusual properties.
We also discuss the time-dependent properties of
these systems, both in the steady state and while approaching it.

\section{Particle-wise disorder}

Consider a system of cars on a single-lane road, too narrow for overtaking to be
possible. Each car has an intrinsic maximum speed, which however may not be 
achieved due to  a slowly moving car in front of it. It is then evident that the slowest car 
in the system will be trailed by a number of intrinsically faster cars, and will
see a larger than average headway in front of it. The question arises:
In the limit of an infinitely long road with a finite density of cars, will 
this headway be finite or infinite? Is there a phase transition between these 
two possible types of behaviour as the density is changed?

This question motivates the study of a collection of particles with a different intrinsic rate of motion for each. In a simple lattice model, studied in
\cite{kf} and \cite{evans}, $N$ particles reside on the sites of a 
one-dimensional ring with $L$ sites. The dynamics involves random sequential
updating of configurations.  An elementary  move consists of an 
attempted rightward hop of a particle, say
the $k$th, with a particle-dependent hopping rate $u(k)$; the move is
actually implemented only if the site immediately to the right is unoccupied. A
configuration is specified by the set of particle locations $\{y_k\}$. One may 
instead consider the set $\{m_k\}$, where $m_k = y_{k+1} - y_k - 1$ is the
headway or gap between the $k$th particle and the one ahead of it.  On a 
periodic ring, a typical set
$\{m_k\}$ represents $N$ equivalent configurations which are translational 
shifts of each other. Evidently the overall density $\rho=N/L$ is 
conserved by the dynamics.

Disorder enters through the selection of hopping rates $u(k)$, each of which 
is drawn independently from a distribution $Prob(u)$.  Especially interesting
is the case with a power law variation near a minimum cutoff speed:
$Prob(u) = [(n+1)/(1-u_o)^{n+1}](u-u_o)^n$ with $u$ in the range
$[u_o,1]$. Here $n$ is an index which characterizes the decay of the
distribution function near the cutoff.  As we shall see below, 
this system shows a phase transition
as the density is decreased, from a state with finite headways ahead of each
particle, to one in which the headway ahead of the slowest 
particle is infinitely long,
or more accurately, a finite fraction of the of the lattice size $L$.

\subsection{Mapping to Zero Range Process}

The analysis of this system is aided by exactly mapping this system with
particle-wise disorder to another system with space-wise disorder 
\cite{kf,evans}. View the
particle index $k$ as labelling urns arranged in a one-dimensional sequence (Fig. 1). The occupancy of urn $k$ is taken to be the length of 
headway $m_k$ in front of the $k$th particle in the original particle 
problem.  Now, a single rightward hop of a
particle reduces the headway in front of it, and augments it behind. In the urn
representation this corresponds to an elementary move of a particle from urn $k$
to urn $k-1$ (Fig. 1).

\begin{figure}[here]
\includegraphics[scale=0.37]{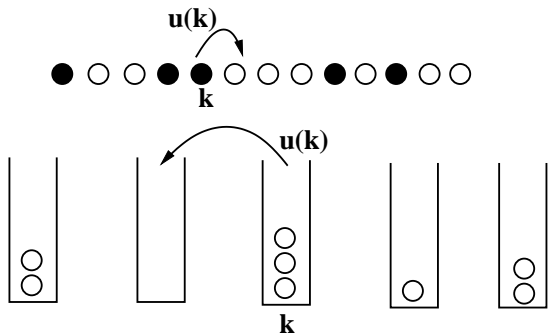}
\caption{Particle-wise disordered model and the Zero-range process}
\label{pwdzrp}
\end{figure}

This model is a special case of the zero range process (ZRP) which describes the
motion of particles between wells or urns, with (in general), occupancy
dependent, site-dependent hopping rates $u(k,m_k)$ \cite{spitzer,eh}.
The ZRP has the virtue that the steady state measure
$P_s(C)$ is known.  In our case, where the rates depend on sites but 
not on occupancies, the steady state weights are particularly simple.
In a grand canonical formulation of the problem we have
\be
P_s(C)= \frac{1}{Z} \prod_{k=1}^{N} \left( \frac{z}{u(k)}
\right)^{m_k} \;. \
\ee
\noindent Here $Z$ is the grand partition function, $m_k$ is the mass 
at site $k$, and $z$ is
the fugacity which is determined by requiring $\sum _k m_k = M$. 
For a given realization of disorder $\{u(k)\}$, the density $\tilde \rho=M/N$
is given by $N^{-1}\sum_k z/(u(k)-z)$. As $\tilde \rho$ is increased, the 
model undergoes a phase transition \cite{benjamini,kf,evans}. 
We use the self-averaging property to replace the sum over $k$ by
an average over $Prob(u)$, separating out the lowest $u$:
\be
\tilde\rho=\frac{1}{N} \; \frac{z}{u_{min}-z}+\int_{u_o}^{1} du \;
\frac{z}{u-z} \;\;f(u) \\.
\ee
where $u_{min}$ is the lowest realized value of $u(k)$; $u_{min}$
approaches $u_o$ in the thermodynamic limit.
The fugacity $z$ increases with 
$\tilde \rho$ until one reaches a critical density $\tilde \rho_c =
u_o(n+1)/n(1-u_o)$  beyond which $z$ gets pinned to $u_o$.  
Beyond this point, the system enters 
a high-density phase in which the excess mass $M-\tilde\rho_c N$ is
`condensed' at the site 
with the lowest hopping rate in the ZRP. The transition is analogous to
Bose-Einstein condensation in the ideal Bose gas or 
a system of bosons in a random potential \cite{luttinger}.

In the particle-wise disordered model,
the condensate represents an infinite headway in front of the slowest car, 
implying an infinitely long jam behind it. Such a jam implies spatial phase 
separation, and exists only for low enough density; at higher densities,
blocking effects between cars reduce the contrast in speeds to such an extent
that headways are interspersed throughout the system and there is no
phase separation.

\subsection {Time-dependent Properties}

It is interesting to ask about the dynamical properties of the system, both 
in the steady state, and also describing relaxation to the steady state.
In steady state, it is evident that the mean current between every 
pair of neighbouring sites in the ZRP is the same. It is less obvious
that the full distribution of time intervals between particle
transfer events is also identical on each site. This property follows from 
the application of Burke's theorem from queueing theory 
\cite{burke,ack} to the problem at hand. This theorem states
that in a reversible birth-death process, a Poisson input of births
results in an identical Poisson distribution of deaths. Identifying the
occupancy of each urn of the ZRP with a population undergoing such a process
then implies the result, as the output from one urn constitutes the
input for the next. In the original particle problem,
this means that the time history of every particle follows an identical 
Poisson distribution despite the fact that each particle has different
intrinsic speeds. This holds in the thermodynamic limit in both the
disordered and the phase separated state \cite{burkebreak}.

Next, let us ask about the behaviour of statistical fluctuations of the
density in the steady state.  In a homogeneous driven system with a 
density-dependent current $\tilde J(\tilde\rho)$, density fluctuations are 
transported around the system as a kinematic wave with
speed $\tilde c = \partial \tilde J/\partial \tilde \rho$, in time 
$N/\tilde c$ \cite{lw,schutzreview}. This wave of fluctuations decays on a time 
scale $\sim N^z (z>1)$, implying that it circulates several times 
around a system with periodic boundary conditions before it dissipates. 
In the disordered ZRP, there are different sorts of behaviour in different 
phases. For $\tilde\rho < \tilde \rho_c$ and for 
$\tilde\rho=\tilde\rho_c,~~ n>1$, the wave speed $\tilde c$ is nonzero.
There is clear evidence of the wave in simulation results for the fluctuations
in the integrated number of particles which move past a site. This
correlation function shows pronounced oscillations with period $N/\tilde c$, 
which ultimately damp down \cite{kjmb}.

In \cite{gtmbprl} it has been argued that quenched disorder does not have a
significant effect on the damping of fluctuations in driven systems if
the kinematic wave speed $\tilde c$ is nonzero. This `kinematic wave
criterion'
implies that there is then {\it no} change in the universality class of the
decay of density fluctuations, which should continue to follow the
behaviour in the absence of disorder. The reason is that in such a case, 
each density fluctuation does not stay long
in the vicinity of a local patch of disorder. To that extent, the effect of
quenched disorder on a ballistically moving density fluctuation  of 
spatial extent $\Delta x$ is akin to temporal noise, and we should 
expect the lifetime of the density fluctuation would continue
to be $\sim(\Delta x)^z$, where $z$ is the dynamical exponent in the absence
of disorder. Monte Carlo
simulations \cite{kjmb} confirm that when $\tilde c \ne 0$, $z$ equals
3/2, the value for a driven system in the absence of disorder \cite{beijeren}.
At $\tilde\rho=\tilde\rho_c$, the speed $\tilde c$ 
vanishes for $n \le 1$. In this case, disorder is expected to have a strong 
effect, and perhaps change the universality class away from that in its absence.
The determination of this class remains an open problem. 

For $\tilde \rho
> \tilde \rho_c$, the system has density $\tilde \rho_c$ in the bulk, with the
surplus mass $(M-\tilde \rho N)$ being on the slowest site.  Thus, 
density fluctuations would be expected to move as a kinematic wave in 
the bulk so long as $n>1$. However, this wave would be absorbed at the 
condensate site and not be able to circulate, implying that correlation
functions in this phase would be non-oscillatory.

Now let us turn to the kinetics of approach to the steady state, starting from a
state with uniform density. In the disordered 
phase $(\tilde \rho < \tilde \rho_c)$, the kinetics is governed by the same
dynamical exponent $z$ that operates in steady state, and the relaxation time
is of order $L^z$, with $z=3/2$. In the condensate phase, however, a
different physical process, namely coarsening,  governs the approach.
Initially, particles hop out of relatively fast sites quite quickly, and
get trapped at slow  sites in the immediate 
neighbourhood. At moderate times, there is thus a finite density
of noticeably large aggregates at such slow sites, which then relax by
releasing their mass to yet slower sites \cite{mbkjpramana}. 
Thus the mass at all such slow
sites, except for the very slowest, shows a nonmonotonic variation in time. 
The mass at the slowest (condensate) site grows as $M_1 \sim t^{\beta}$,
with $\beta=(n+1)/(n+2)$, a result first derived for a model with
deterministic dynamics \cite{newell}, but which remains true 
even in the stochastic model under consideration \cite{krug,kf,kjmb}.
This translates into a statement about the growth of headway lengths $\xi$
in the particle model
\be
\xi(t) \sim t^{\frac{n+1}{n+2}}.
\label{xi1}
\ee
A similar coarsening description for headways is also found 
to hold in a lattice gas
model of traffic, which incorporates features of acceleration and slowing
down, and in which different drivers have different propensities for random
braking \cite{chowdhury}.

Furthermore, the problem has also been studied using
open boundary conditions, with injection and ejection rates specified
\cite{bengrine1}.
In the absence of disorder, it is known that the open system
exhibits phase transitions between a low-density phase, a high-density phase,
and a high-current phase, as the injection and ejection rates are varied
\cite{derrida,schutzdomany}. With particle-wise disorder, 
there is a  shift of the
phase boundaries owing to the breaking of particle-hole symmetry. Within
the low-density phase, there is a regime in which the drift velocity of
all particles is determined by the slowest \cite{bengrine1}, as with 
perodic boundary conditions.

Finally, we discuss some recent work on two-way traffic \cite{igloi1,igloi2}. 
Consider particle-wise disorder in which the majority of particles move 
preferentially rightward, while a minority (a fraction $f$) move peferentially
leftward. For each particle, the ratio of the larger hopping rate to the
smaller one, which is a measure of bias, is taken to be fixed, 
but the direction of hopping is
random. The problem was mapped onto a ZRP, which inherits the random
easy-direction property. This system has much in common with the
spatially disordered ASEP with bidirectional bonds, to be discussed
in the next section. In particular, as in that case, the current
decays as an inverse power of the size, with the power depending continuously
on both $f$ and the bias.

\section{Space-wise disorder}

Consider the transport of interacting particles which are driven through
a randomly disordered one-dimensional path. 
To address the question of what sort of macroscopic states result, we restrict
ourselves to the disordered asymmetric simple exclusion process. 
Disorder enters through the transition rates, which are 
assigned randomly to the bonds between neighbouring sites, e.g. 
$u_{i,i+1}$ describes the rate of attempted hopping
from site $i$ to $i+1$, while $u_{i+1,i}$ is the attempt rate in the reverse
direction. A hop to a site actually occurs only if the site in question is
unoccupied. Results obtained for a particular disorder realization $R \equiv
\{u_{i,i+1},u_{i+1,i}\}$ are then to be averaged over $R$ using a
pre-specified probability distribution, taken to be the product of 
identical and independent distributions across all bonds. The larger
of the two rates on each bond defines the easy direction 
on that bond. Below we will
distinguish between two situations: {\it unidirectional}, in which
the easy direction of every bond is the same, but the strength
is a random variable ((a) and (b) in Fig. 2) and {\it bidirectional},
in which the easy direction is itself a random variable ((c) and (d)
in Fig. 2).

In the unidirectional case, we will further specialize to only the case of no 
backward hopping $(u_{i+1,i}=0)$:
\be
Prob(u_{i,i+1})= (1-f)\delta(u_{i,i+1}-u_1) + f\delta(u_{i,i+1}-u_2)
\ee
\noindent with $u_2<u_1$. The fraction $f$ and relative strength $r=u_2/u_1$
of weak bonds specify the extent of disorder.

In the bidirectional case, the easy direction
of each bond is taken to vary randomly, but the relative hopping rate $\lambda$
against and along the easy direction on each bond is taken to be the same.
\be 
Prob(b_{i,i+1} =\frac{u_{i+1,i}}{u_{i,i+1}})=(1-f)\delta(b_{i,i+1}-\lambda) +
f\delta(b_{i,i+1} - \lambda^{-1}).
\ee
Here $f$ is the fraction of backward pointing bonds.

The large scale behaviours that result depend on the type of disorder
as well as on the overall density $\rho$.  Figure 2 shows a schematic depiction
of a segment of the disordered lattice, the corresponding variation 
of the current
$J$ with system size $L$, and the schematic spatial variation of the
mean occupancies $\rho_i \equiv \langle n_i \rangle$ in a particular realization 
of disorder in different cases. These are discussed below.

(a) With unidirectional disorder and a particle density that is sufficiently
far from $1/2$, i.e. $ |\rho-1/2| > \Delta$, the density is homogeneous
on the macroscopic scale. The value of $\Delta$ depends on the concentration
$f$ and relative strength $r=u_2/u_1$ of the weak bonds. The current $J$ 
approaches a nonzero value in the thermodynamic limit. 

(b) With unidirectional disorder and $\rho$ close to $1/2$, i.e.
$ |\rho - 1/2| < \Delta$, the state exhibits phase separation;
the system is characterized by two distinct values of the density,
each extending over a macroscopic region.  As in (a), $J$ has a finite
value in the thermodynamic limit. This value of the current is the same
for all $\rho$ in the regime $| \rho - 1/2| < \Delta$.

\begin{figure}[here]
\includegraphics[scale=0.4]{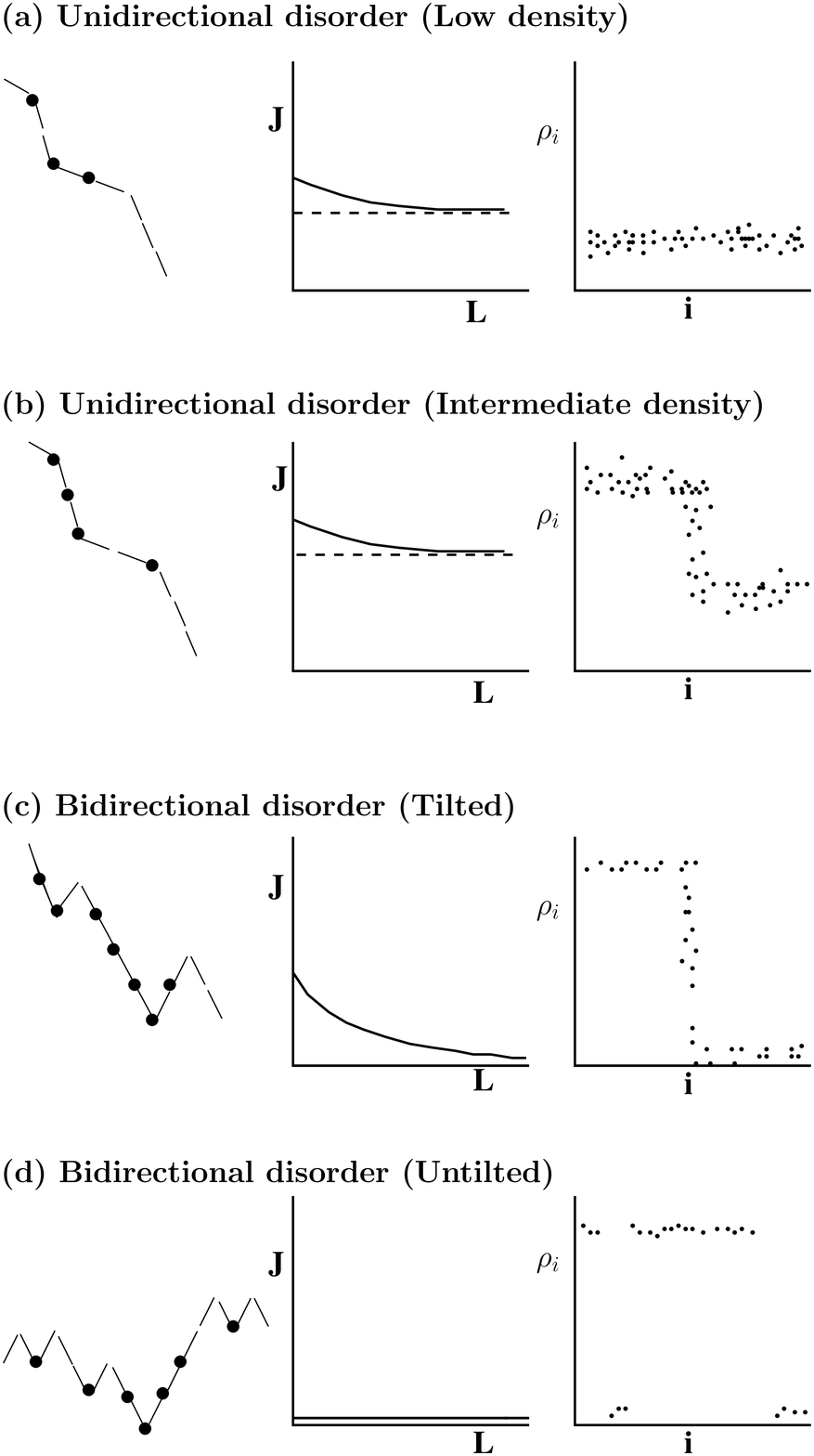}
\caption{Space-wise disorder: depiction of possible behaviours}
\label{swd}
\end{figure} 

(c) With bidirectional disorder, but with an asymmetry in the number of
forward and backward bonds $(f \ne 1/2)$, the potential has local minima
and an overall tilt.  On a macroscopic scale, the system has two distinct values
of the density, close to 1 in one region and close to 0 in the other. The
current $J$ falls with increasing system size $L$ as an inverse power of $L$.

(d) With bidirectional disorder and equal numbers of forward and backward
bonds $(f=1/2)$, the potential has local minima, but no overall tilt. 
Consequently the mean current is zero. The density profile shows large
sample-to-sample variations, but every sample has a high-density region 
that extends over a finite fraction of the system size.

Let us turn to an explanation of the features described above. Central to the
discussion is the observation that while there are strong spatial variations 
in the local density $\langle n_i \rangle = \rho_i$, the steady state value of the current
$J$ is uniform across all bonds.

\subsection {Unidirectional disorder} 

 The time averaged current in bond $(i,i+1)$ is
$J_{i,i+1} = u_{i,i+1} \langle n_i(1-n_{i+1} ) \rangle$.
Since the exact steady state
weights of particle configurations $\{ n_i\}$ are not known,
one cannot compute the two-point correlation function involved. To
proceed, we employ a mean field approximation which ignores
correlations betwen site densities in the steady state, and replaces
$\langle n_i n_j \rangle$ by 
$\langle n_i \rangle \langle n_j \rangle = \rho_i \rho_j$, but does allow for spatial variation
of $\rho_i$, which is crucial. To find the mean field solution $\{ \rho_i \}$
we may proceed in two ways. (i) For a given value of $J$, iterate the
set of equations $\rho_{i+1}=1-J/u_{i,i+1}\rho_i$ around the 
periodic chain, until convergence is achieved. (ii) For a given value
of the overall density, evolve local densities (in fictitious time) through 
$\rho_i(t+1)=\rho_i(t)+j_{i-1,i}(t)-j_{i,i+1}(t)$  where the
instantaneous currents are given by $j_{i,i+1}(t)=u_{i,i+1}\rho_i(t)
[1-\rho_{i+1}(t)]$. The densities then converge to time-invariant values,
and the corresponding current is uniform across bonds. As discussed in
\cite{gtmbpre}, method (ii) is
preferable, as values of $J$ above a threshold value are not allowed,
and method (i) does not converge well near the threshold. The resulting
mean field density profile can be tested against the results of direct 
Monte Carlo simulation. The agreement is remarkably good, in that the mean field
approximation tracks every one of a macroscopic number of `minishocks'
and correctly predicts their positions, though it does not 
accurately reproduce their shapes \cite{gtmbpre}.

\begin{figure}
\includegraphics[scale=0.5]{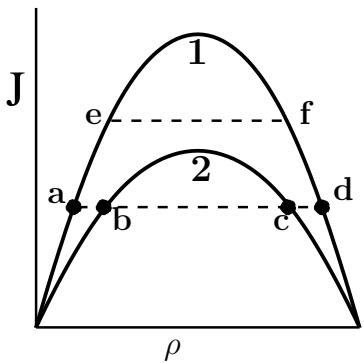}
\caption{Current-density plots}
\label{jrho}
\end{figure}   

The principal result is the existence of a density range $\Delta$ in
which the current has a constant, density-independent value (regime (b)),
and where the system exhibits phase separation of densities \cite{gtmbpre}. 
A simple
qualitative understanding of both these features can be obtained by referring
to Fig. 3. The argument relies on a conjectured maximum current principle,
which states that for a given value of the overall density, the system
settles into a state which maximizes the mean current. In our system,
let us suppose that the density in each stretch of like bonds is uniform,
and label  by 1 and 2 the stretches of strong and weak bonds respectively. 
The two parabolas in Fig. 3 are the corresponding $J$ versus $\rho$ curves
for the pure 1 and pure 2 systems. Since the current is uniform in the
disordered system, a value of current such as shown by by the upper
dashed line $ef$  in the figure
is ruled out, as such a value cannot be sustained in a 2-stretch. By contrast,
a value of $J$ corresponding to say $ad$ is quite possible, as the 
(1,2) densities can assume the values $(\rho_a,\rho_b)$ if the 
density is relatively low, or alternatively $(\rho_c,\rho_d)$ if the
density is relatively high. Such choices correspond to the regime
$|\rho-1/2|>\Delta$, and result in a density profile that is homogeneous
on a macroscopic scale as strong and weak bonds are interspersed
at the microscopic level. If the density is too close to half-filling 
$|\rho - 1/2|< \Delta$, the system adjusts by keeping the current
constant at its maximum allowed value, and replacing
$(\rho_a,\rho_b)$ in (1,2) stretches by  $(\rho_c,\rho_d)$, in a  finite fraction
of the full lattice. The resulting state then shows phase separation
of the density. The location of the high density region is decided 
by the longest stretch of weak bonds in the system, as this stretch
(whose length is of order ${\rm} lnL$) mimics the effect of a single 
defect weak bond in an otherwise pure system, which is known to 
induce phase separation in a finite range of density \cite{janowsky,schutz}.

Interestingly, it is possible to obtain upper and lower bounds \cite{krug}
on the value of the threshold density for phase separation
$\rho_c= 1/2-\Delta$ in terms of the ratio of bond strengths $r=u_2/u_1<1$
and the fraction $f$ of weak bonds. An upper bound is obtained on noting that 
the maximum current that can flow through a stretch of slow bonds decreases
as the stretch length increases. It is then plausible that the current would
be the least in the fully segregated limit where all slow bonds form a
single large stretch \cite{gtmbpre}. In such a system consisting of two 
connected homogeneous parts, it is easy to figure out the way a given 
overall density would distribute itself over the two portions. A lower
bound on $\rho_c$ was obtained \cite{krug} by arguing that the 
current in the disordered
ASEP is bounded above by the current in a corresponding ZRP with the
same disorder distribution and an equal number of particles.


The results discussed above hold for the case of binary disorder in the 
transition rates. Similar phenomena, e.g. the plateau in the $J-\rho$
curve over a finite range $\Delta$ of density, have been found also 
for more general, continuous distributions of the transition rates
\cite{bengrine2,harris}. The problem has also been studied using
open boundary conditions, specifying injection and ejection rates.
An argument based on moving shock fronts \cite{popkov} suggests that in the 
presence of disorder, the high-current phase extends over a larger region of 
the phase diagram \cite{harris} than in the pure system \cite{derrida}. 
Further, a numerical study of the first order
transition between the low and high density phases shows that the transition
remains first order, but the location of the transition shows strong sample 
to sample fluctuations which do not seem to damp down in the thermodynamic 
limit \cite{enaud}.

Going beyond the ASEP, the effect of slow sites on a lattice model of traffic
was studied by numerical simulation in \cite{vicsek}. Interestingly, a
regime with two current plateau regions was observed, and the system was
observed to switch between the two branches.

To conclude this section, we mention an intriguing symmetry that  holds for
the disordered ASEP when no backward hopping is allowed. A typical
configuration of quenched disorder is not symmetric under space inversion,
and consequently nor is the steady state density profile. In spite of this,
the magnitude of the steady state current was observed to be invariant
under reflection \cite{gtmbpre}. Subsequently, this invariance property
has been proved for both periodic \cite{goldstein,punnoose} and open 
\cite{goldstein} systems.

\subsection{Bidirectional disorder: With tilt}

When the easy direction on a bond is a random variable, one needs to distinguish
between the cases $f<1/2$ and $f=1/2$, corresponding to a tilted and 
untilted potential respectively (figs. 2(c) and (d)). 
We may imagine assigning arrows to denote the easy direction
on the bonds of a periodic ring; a fraction $(1-f)$ point rightward, while
a fraction $f$ point leftward. A particle-hole exchange occurs in the direction
of the arrow with rate $u$, and in the opposite direction
with rate $u\lambda$ with $\lambda<1$.

In the tilted case, there is an overall tendency for the particles to move
rightward, but the question is whether there is a nonzero current in the
thermodynamic limit. Local barriers to the rightward motion occur in
the form of backbends which consist of a chance agglomeration of
successive left-pointing arrows. An upper bound $J_l$ on the current
that can be carried by a backbend of length $l$ is obtained by
considering the boundary conditions, $\rho=1$ at the leftmost
end of the backbend and $\rho=0$ at the rightmost end, which
force the largest possible current through it Within a
mean field approximation, the density profile, and hence the current,
can be calculated \cite{rrmb} with the result
 $J_l \sim \lambda^{\frac{1}{2}l}$. The factor $\frac{1}{2}$ is a consequence of
particle-hole symmetry which implies that the particle density extends up to
half-way through the stretch, so that the topmost particle needs to be 
activated across ${\frac{1}{2}l}$ sites.
An exact calculation \cite{blythe} confirms the exponential decay of current
with $l$. The current in the full system of size $L$
 is limited by the length $l^*(L)$
of the longest backbend. Since the probability of the occurrence of
$l$ successive backward bonds is $f^l$, we may estimate $l^*$ from
$Lf^{l^*} \simeq O(1)$. The estimate of the current $\lambda^{l^*/2}$
can be written as
\be
J(L) \sim L^{-\theta/2}
\label{current}
\ee
where the exponent  
\be
\theta= {\rm ln} \lambda/{\rm ln} f 
\label{theta}
\ee
depends explicitly on the bias and
fraction of reverse bonds. This dependence has been confirmed by Monte Carlo 
simulations \cite{gtmbpre}. 

\subsection{Bidirectional disorder: Untilted} 

When there is no overall tilt, the mean current is identically zero.
The problem then reduces to the equilibrium problem of hard core 
particles at temperature $T$ in a static random  potential 
corresponding to a  height profile $\{h_i\}$. The Hamiltonian is
\be
{\cal H}= \sum_i h_i n_i.
\ee
Height differences are related  to the hopping rates of the model
through the condition of detailed balance
\be
exp[-(h_{i+1}-h_i)/T]=\frac{u_{i,i+1}}{u_{i+1,i}}.
\ee
Since each local slope of the height
is equal in magnitude but random in sign, the full height profile
is isomorphic to the trail of a random walker, with the height being the 
displacement of the walker and $i$ being time.  Thus the problem reduces to 
that of hard-core particles in a Sinai potential.
In the limit of large bias ($T \rightarrow 0$), 
the system approaches the ground state, which can be found as 
follows. Make a constant-height cut of the height profile, fill particles
up to that point, and then raise the level of the cut step by step
and fill in particles till they are exhausted. 
The lengths $\ell$ of particle clusters (which are present below the 
topmost filled level) are then determined by returns to the 
origin of the walker, and
their distribution follows an asymptotic power law $Prob(\ell) \sim
\ell^{-3/2}$. This distribution has an infinite mean, and in a system of
size $L$ an estimate based on extremal statistics shows that the largest 
cluster size is $gL$. The prefactor $g$ shows strong fluctuations from
one sample to another, but is always of order unity. Thus a typical 
configuration consists of at least one $O(L)$-sized cluster in addition to
$\sim \sqrt{L} $ clusters of size $\sim \sqrt{L}$. 

In \cite{dd}, a
similar model (the Coarse-grained Depth or CD model) was analysed, and 
the disorder-averaged two-point correlation 
function $C_L(x) \equiv \langle n_i n_{i+x} \rangle _L-\rho^2$ 
was computed, using the cluster size
distribution $Prob(\ell)$ as input. The result is that $C_L(x)$ is a scaling
function of argument $x/L$:
\be
C_L(x) = Y({{x}/{L}}) \approx m_o^2[1-a({x}/{L})^{\alpha}],~~~~~x/L<<1
\label{cusp}
\ee
with intercept $m_o=1$ and $\alpha=1/2$.

The argument of the scaling function indicates that it describes a 
phase-separated state \cite{bray}. The finite intercept $m_o^2$ is a measure of 
long-range order. The cusp singularity is a manifestation  of ill-defined 
interfaces, as it is known that a phase-separated system with well defined
interfaces has $\alpha=1$, in concordance with the Porod law \cite{bray}.
We conclude that the state exhibits an unusual sort of phase separation,
with unusual interfacial characteristics and strong sample to sample 
fluctuations.

\subsection{Time-dependent properties}

In this subsection we discuss both the steady state dynamics and the coarsening
dynamics of systems with unidirectional and bidirectional disorder.

With unidirectional disorder, the behaviour of density 
fluctuations in steady state resembles that in the zero-range process
discussed in the previous section. For $|\rho-\rho_c|>\Delta$, there 
is a kinematic wave which circulates through the system, and induces
long-period oscillations in the density \cite{gtmbprl}. The dissipation 
was monitored by performing a Galilean shift to keep up with the wave.
As expected on the basis of the kinematic wave criterion discussed earlier,
disorder was found to be irrelevant, and $z=\frac{3}{2}$ still holds
\cite{gtmbprl}.  For $\rho=\frac{1}{2}$, on the other hand the system is
phase separated, and the system supports two kinematic waves which
move in oppsite directions in the two coexisting phases. Simulations
suggest that the kinematic waves emanate from one interface, and
terminate at another, leading to a different universality class
\cite{gtmbprl}.

In the phase-separated regime, the kinetics of approach to the steady state
involves a coarsening process \cite{krug}. Large-density stretches accumulate
behind bottleneck regions which are stretches of consecutive weak bonds. 
If $\Delta J$ is the difference of currents exiting from two such bottlenecks
separated by a distance $\xi$, the region in between would fill up in 
time $t\sim \xi/\Delta J$. Estimating the probability of occurrence
of stretches of length $l$ and recalling that they carry a current
$J_l=u_2/4(1+a/l)$ \cite{krugmeakin} leads to the estimate
\be
\xi (t) \sim \frac{t}{ln t}.
\label{xi2}
\ee

Turning to bidirectional disorder with tilt ($f<\frac{1}{2})$, recall that
in steady state, particles are held back behind the longest backbend,
while the region ahead of it is empty. The coarsening
properties of this model were obtained by Krug \cite{krug} who pointed out that
bottleneck stretch lengths are much smaller than their separations,
implying that we may mentally replace stretches by single bonds that
allow a maximum current $J_l\sim\lambda^{{\frac{1}{2}}l}$ to flow
through. Following through, we find that the coarsening length is given by
\cite{krug}
\be
\xi(t) \sim t^{\frac{1}{1+\theta/2}}.
\label{xi3}
\ee

In the untilted case, the steady state is the equilbrium state of hard
core particles in a random (Sinai) potential. Not much is known about
the dynamics in this case, though the single particle problem
is well studied \cite{bouchaud}. The existence of large $(\sim O(\sqrt L)$
 potential barriers would make the approach to equilibrium logarithmically
slow, and it would be interesting to see how much slower the
coarsening process
is than the familiar diffusion of a single particle in a Sinai landscape.

\section{Dynamical disorder}

In this section, we study the properties
of an ASEP subjected to bidirectional bond disorder, where the disorder
is itself time-dependent. Specifically, we consider an untilted 
potential (as in Fig. 2(d)) which is evolving stochastically in time through
the single-step model \cite{plischke} which preserves 
the overall no-tilt condition under time evolution.
The basic dynamical move is the  interchange of nearest
neighbour  links of the height field 
at a certain rate if they represent
opposite slopes. If $L$ and $R$ stand for left-leading and right-leading 
links respectively, then a link configuration $....LR....$ can evolve
into $....RL....$ at a certain rate, and the reverse move is allowed at
an equal rate. The large space and time properties of the 
height field of this model are described by the continuum
Edwards-Wilkinson model \cite{EW}. The
ASEP  particles are passive sliders on the surface, in that they are
influenced  by, but do not back-influence, the dynamics of
the fluctuating surface.

This problem has been studied in \cite{dd}. 
Particles tend to form clusters when they fall into valleys, and to decluster
when valleys overturn and form hills. However, declustering involves
activation and is a slow process, while clustering is relatively rapid.
This asymmetry leads to overall clustering. Numerical simulations show
that this tendency to cluster persists at all length scales, and leads
finally to a steady state that exhibits
phase separation. However, this phase separated state is unusual in that
is characterized by large fluctuations of the order parameter,
which do not damp down in the thermodynamic limit. Despite the
presence of strong fluctuations, the system never loses its ordered
character \cite{scmb}. Fluctuations carry the system relatively
rapidly from one ordered configuration to another macroscopically distinct  one.
But the probability for the system to leave this
attractor of ordered states vanishes exponentially in the system size.

A quantitative measure of the ordering is provided by the two point
correlation function. Results of simulations show that Eq. \ref{cusp},
derived for the disordered problem with an untilted, bidirectional
potential describes the form of $C_L(x)$ 
in this case as well. When the update rates for the landscape and particles
are equal, the value of $m_o^2$ is close to 0.8, indicating
long-range order, consistent with the existence of a  phase ordered state.
Moreover, the value of the cusp exponent $\alpha$ is close to
$\frac{1}{2}$, its value for the problem discussed in the previous section.
This quantitative similarity between the scaling properties
of phase ordering induced by quenched disorder, and by dynamical disorder
corresponding to an evolving Edwards-Wilkinson landscape, 
is not completely understood.

\subsection{Time-dependent properties}

The time dependence of the particles is dictated by fluctuations of the driving
height field. Except at very small times, the autocorrelation function
$A_L(t)\equiv \langle n_i(0)n_i(t) \rangle $ is found numerically\cite{scmb} to be a scaling function
\be
A_L(t) = X(t/L^z) \approx m_o^2[1-d({t}/{L^z})^{\beta}],~~~~~t/L^z<<1
\label{cuspdynamics}
\ee
with $z=2$ being the dynamical exponent governing the height-field dynamics
and $d$ a constant of order unity.  Simulations show that
 $m_o^2\simeq0.8, \beta \simeq 0.2$. Like its
counterpart for static correlations (Eq. \ref{cusp}), the scaling function $X$
shows an approach with a cusp singularity to a finite value at 
small argument, once again
signalling long range order of an unusual sort. Finally, to understand
the coarsening properties of this model, recall that clustering occurs by
collecting particles in valleys of the driving height field. Since
the  largest valleys that form in time $t$ are typically of size $t^{1/z}$,
we expect the coarsening length in this case to follow
\be
\xi(t) \sim t^{\frac{1}{2}}.
\label{xi4}
\ee
The time-evolving two-point correlation function 
$ \langle n_i(t)n_{i+x}(t) \rangle-\rho^2$ 
is expected to be similar to
the form of Eq. \ref{cusp}, except that the system size $L$
is replaced by the coarsening length $\xi(t)$. 
The results of numerical simulations agree well with this \cite{dd}.

\section{Conclusion}

We have reviewed the properties of the one-dimensional ASEP with various
types of disorder --- particle-wise, space-wise (unidirectional
and bidirectional), and dynamical.  All forms
of disorder bring about phase separation, though under different
conditions, and of different types.

With particle-wise disorder, phase separation occurs below a critical density 
$\rho_c=(1+\tilde \rho_c)^{-1}$. The macroscopic stretch ahead of the slowest
particle is totally devoid of particles $(\rho=0)$, whereas the stretch
behind is mixed and has $\rho=\rho_c$.

With unidirectional space-wise disorder, the system supports phase separation
in the region $\rho_c<\rho<1-\rho_c$, with coexisting phases that have densities
$\rho_c$ and $(1-\rho_c)$. 

With bidirectional disorder with tilt, the system phase separates at all values
of $\rho$ into phases with densities close to 0 and 1. The deviations from
these values are proportional to the value of the current (Eq. \ref{current}),
and approach zero as $L \rightarrow \infty$. 

With bidirectional disorder and no tilt, we have the problem of hard core
particles in a Sinai potential. The phases are again characterized by 
$\rho=0$ or $1$, as evidenced by $m_o=1$ in Eq. \ref{cusp}, but have
some unusual properties: the interfacial
region is very broad, and there are strong sample to sample variations in
the length of the longest ordered stretch.

Finally, the state reached through dynamical disorder is surprisingly similar
to the previous case, with the difference that  $m_o \ne 1$ in this case.

Turning to dynamical properties in the steady state, we have seen that 
results are consistent
with the kinematic wave criterion \cite{gtmbprl}, namely that a finite velocity
of the  kinematic wave renders disorder irrelevant;  
the universality class characterizing the decay of fluctuations remains
unchanged. The determination of the disorder-determined
universality class when the wave
speed vanishes remains an open problem, for instance, 
for particle-wise disorder with $n\le 1$. 

The coarsening behaviour characterizing the approach to the steady state is 
well understood in several cases (Eqs. \ref{xi1}, \ref{xi2}, 
\ref{xi3}, \ref{xi4}).  However, the coarsening properties
of a system of hard core particles in the Sinai potential, which corresponds
to the untilted bidirectional case, remains an interesting open problem.
\vspace{0.5cm}
\section*{Acknowledgements}

I thank Goutam Tripathy, Dibyendu Das, Kavita Jain, Sakuntala Chatterjee 
and Satya Majumdar for collaboration on several of the topics discussed 
above. I thank Satya Majumdar 
and Gunter Sch\"utz for useful comments on this paper, and Shamik Gupta for 
help with the figures.

I am grateful to the Isaac Newton Institute, Cambridge for warm hospitality 
and  acknowledge support through ESPRC grant 531174.
I also acknowledge the support of the Indo-French Centre for the Promotion of
Advanced Research (IFCPAR) under Project 3404-2.

\end{document}